\documentclass[aps, pra, reprint, twocolumn, flushbottom, superscriptaddress, floatfix]{revtex4-1}
\usepackage{graphicx}
\usepackage{braket}
\usepackage{amsmath, amssymb}

\newcommand{\pref}{P_R}
\newcommand{\psam}{P_S}
\newcommand{\ploss}{P_L}
\newcommand{\qref}{\Ket{R}}
\newcommand{\qsam}{\Ket{S}}
\newcommand{\qloss}{\Ket{L}}
\newcommand{\dal}{\Delta \alpha}
\newcommand{\dpr}{\Delta P}
\newcommand{\nmin}{n_\mathrm{min}}
\newcommand{\nloss}{n_\mathrm{loss}}
\newcommand{\nopt}{N_\mathrm{opt}}
\newcommand*\mean[1]{\overline{#1}}

\begin{document}

\title{Semitransparency in interaction-free measurements}

\author{Sebastian Thomas}
\email{sebastian.thomas@fau.de}
\affiliation{Friedrich-Alexander University Erlangen-N\"urnberg, Department of Physics, 91058 Erlangen, Germany}
\affiliation{Max-Planck-Institut f\"ur Quantenoptik, 85748 Garching bei M\"unchen, Germany}
\author{Christoph Kohstall}
\affiliation{Physics Department, Stanford University, Stanford, California 94305, United States}
\author{Pieter Kruit}
\affiliation{Department of Imaging Science and Technology, Faculty of Applied Sciences, Delft University of Technology, 2628~CJ Delft, The Netherlands}
\author{Peter Hommelhoff}
\affiliation{Friedrich-Alexander University Erlangen-N\"urnberg, Department of Physics, 91058 Erlangen, Germany}
\affiliation{Max-Planck-Institut f\"ur Quantenoptik, 85748 Garching bei M\"unchen, Germany}

\date{\today}

\begin{abstract}
We discuss the effect of semitransparency in a quantum-Zeno-like interaction-free measurement setup, a quantum-physics based approach that might significantly reduce sample damage in imaging and microscopy. With an emphasis on applications in electron microscopy, we simulate the behavior of probe particles in an interaction-free measurement setup with semitransparent samples, and we show that the transparency of a sample can be measured in such a setup. However, such a measurement is not possible without losing (i.e., absorbing or scattering) probe particles in general, which causes sample damage. We show how the amount of lost particles can be minimized by adjusting the number of round trips through the setup, and we explicitly calculate the amount of lost particles in measurements which either aim at distinguishing two transparencies or at measuring an unknown transparency precisely. We also discuss the effect of the sample causing phase shifts in interaction-free measurements. Comparing the resulting loss of probe particles with a classical measurement of transparency, we find that interaction-free measurements only provide a benefit in two cases: first, if two semitransparent samples with a high contrast are to be distinguished, interaction-free measurements lose less particles than classical measurements by a factor that increases with the contrast. This implies that interaction-free measurements with zero loss are possible if one of the samples is perfectly transparent. A second case where interaction-free measurements outperform classical measurements is if three conditions are met: the particle source exhibits Poissonian number statistics, the number of lost particles cannot be measured, and the transparency is larger than approximately $1/2$. In all other cases, interaction-free measurements lose as many probe particles as classical measurements or more. Aside from imaging of gray levels, another possible application for interaction-free measurements is the detection of arbitrarily small phase shifts in transparent samples.
\end{abstract}

\maketitle

\section{Introduction}
In some applications of imaging and microscopy, the damage that is inflicted on a sample while its image is taken is the main limit on what kind of samples \emph{can} be imaged. Particularly in electron microscopy, the large radiation dose that any sample receives can make the imaging of, e.g., living biological samples impossible~\cite{Spence2013, Egerton2004}. Hence, reducing sample damage is crucial for future developments of electron microscopy. Next to other proposals~\cite{Okamoto2012}, a quantum mechanical protocol called ``interaction-free measurement'' (IFM), previously proven to work with photons~\cite{Elitzur1993, Kwiat1995}, has been proposed as a means to this end~\cite{Putnam2009}. The basic idea of interaction-free measurements is to exploit the wave-like features of quantum particles in order to gain information about an object while reducing the interaction between particle and object to a minimum. This is accomplished by confining the probe particle in a resonator in which it makes multiple round trips. During each round trip, a small part of the wave (the ``sample wave'') is split off from the original trajectory (the ``reference wave'') and sent through the sample. After many round trips, the presence of a sample can be inferred from the intensity of the reference wave even though the total intensity in the sample wave has been arbitrarily small. This is explained in detail in section~\ref{ifm}.

The field of interaction-free measurements started with a discussion on ``negative-result'' measurements, where the location of an object is inferred from \emph{not} being measured with a detector~\cite{Renninger1960}, which leads to a change in the wave function of the object~\cite{Dicke1981}. Elitzur and Vaidman proposed an IFM scheme employing a Mach-Zehnder interferometer which sometimes detects an absorbing object without any absorption occurring~\cite{Elitzur1993, Kwiat1995}. In this simple interferometric scheme, only some measurement runs constitute a successful IFM while the probe particle is absorbed in the other runs. The rate of successful IFM runs can be increased arbitrarily close to $1$ in more elaborate interferometric setups with multiple round trips through the path containing the sample~\cite{Kwiat1995, Kwiat1999}.

All applications of interaction-free measurements may come with imperfect absorbers like semitransparent objects. Previous work on semitransparency in IFMs has shown that the rate of successful IFM runs is reduced if a semitransparent object is to be detected instead of a perfect absorber~\cite{Jang1999, Vaidman2002, Garcia-Escartin2005}. This can be compensated by increasing the number of round trips in the interferometer~\cite{Kwiat1998, Azuma2006}. In this article, we study the effect of semitransparent samples in IFM setups, and we calculate the damage that arises during either an IFM or a conventional measurement in two different situations relevant for imaging: (1) discriminating between two objects which have different transparencies or (2) determining the transparency of an object. We also compare our results to a lower bound for the damage in general quantum measurements of semitransparency, which was derived by Massar, Mitchison, and Pironio for a generalization of interaction-free measurements~\cite{Mitchison2001, Massar2001, Mitchison2002}. Additionally, we discuss the effect of phase shifts in IFMs.

Most of the previous discussion of interaction-free measurements has been focused on the detection of absorbing objects using photons as probe particles. Hence, such measurement schemes have sometimes also been called ``absorption-free'' measurements~\cite{Mitchison2001}. Especially if other probe particles like electrons~\cite{Putnam2009} or neutrons~\cite{Hafner1997} are considered, however, it should be noted that the IFMs are not only absorption-free but also free of any process that prevents the probe particle's wavefunction from continuing undisturbed on its original path. Scattering out of the path or momentum-changing collisions turn out to have the same effect as absorption. Therefore, we will call all these processes ``loss'', as in ``lost for further interaction with the reference wave''. Many of the interaction processes, such as inelastic scattering or electron knock-out processes, cause damage to the sample, which can be prevented using IFM schemes. While the relationship between loss and damage is complicated and depends on the specific setup and sample, we will assume in the following that a higher probability of ``loss'' means more damage. Interaction-free measurements open up novel applications in cases where sample damage is particularly disruptive. Next to electron microscopy~\cite{Putnam2009}, other potential applications are the imaging of photo-sensitive materials~\cite{Inoue2000} or single-atom detection~\cite{Karlsson1998, Volz2011}.

\begin{figure}[t]
\includegraphics[width=.9\columnwidth]{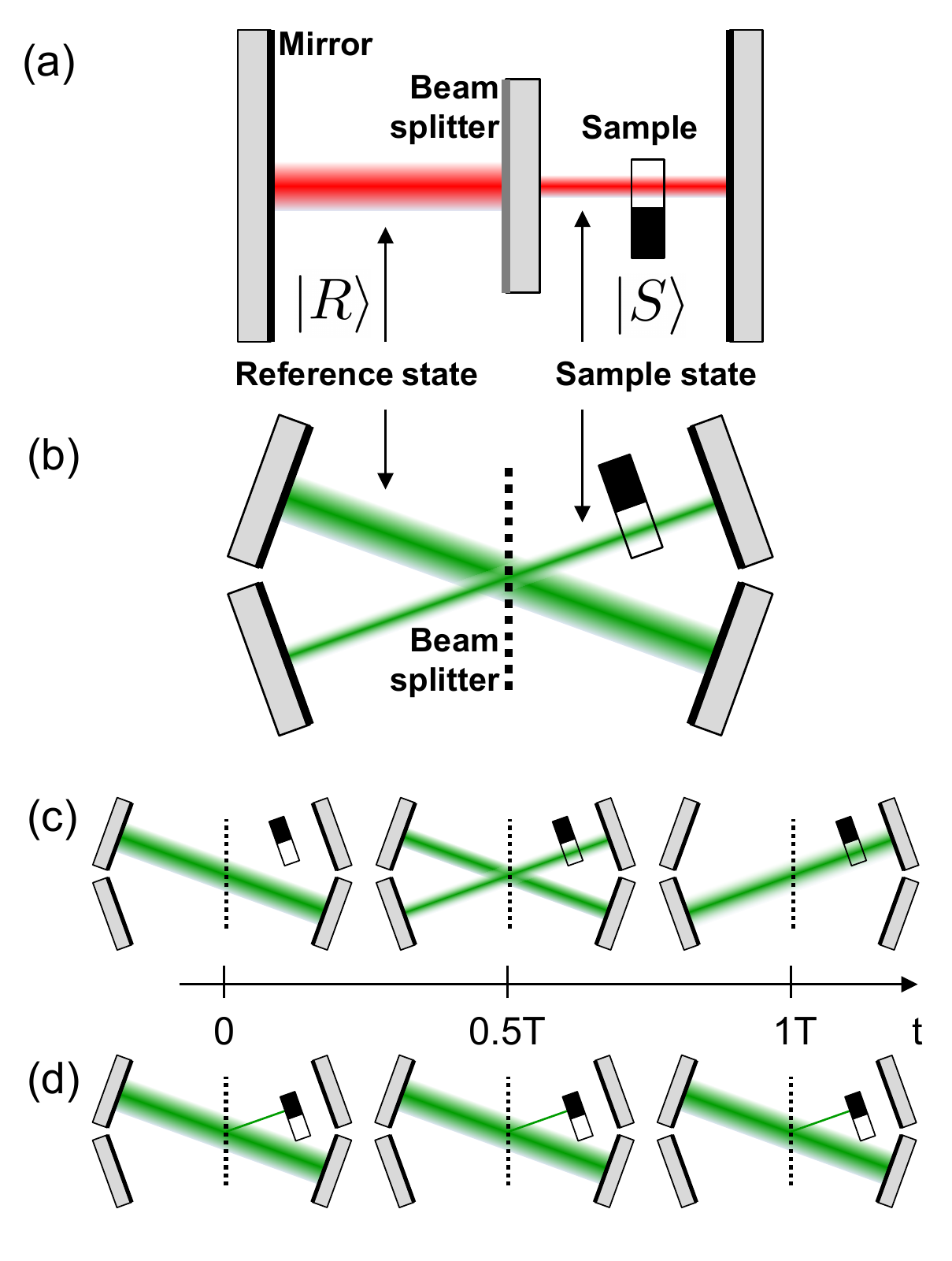}
\caption{Sketches of quantum-Zeno-like IFM setups. (a) The light mode in the left reference cavity is coupled via a semi-transparent mirror to a mode in the sample cavity on the right, as proposed in Ref.~\cite{Kwiat1995}. Figure~(b) shows a possible realization of two coupled cavities for electrons with a diffraction-based beam splitter. (c) With a transparent sample in the sample beam, the probability of finding the electron will coherently oscillate between the reference and sample beam. (d) With a lossy sample in the sample beam, the coherent build-up of probability amplitude in the sample beam is prevented and the electron stays in the reference beam. Depending on whether we measure the probe particle in the sample or in the reference beam, we determine to have a transparent or lossy sample, respectively, in the sample beam. The same concept as in (c) and (d) applies to (a).}
\label{sketch}
\end{figure}

\begin{figure}[tb]
\includegraphics[width=\columnwidth]{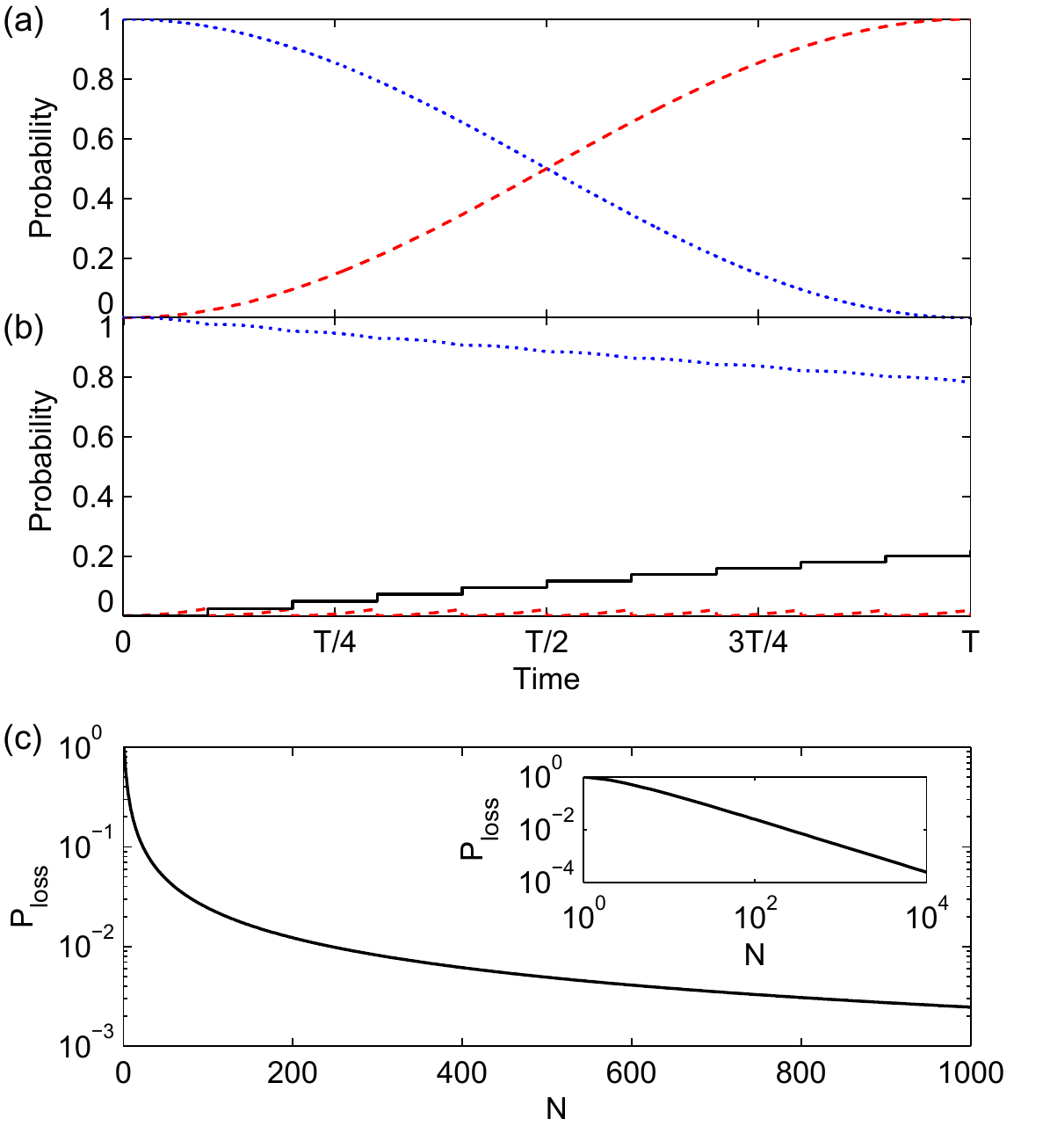}
\caption{Probabilities $\pref$, $\psam$, and $\ploss$ of finding the probe particle in the reference state $\qref$ (dotted blue line), sample state $\qsam$ (dashed red line), or loss state $\qloss$ (black line). (a) Probabilities versus time for a perfectly transparent sample, as in Fig.~\ref{sketch}~(c). At time $T$ the particle is in the sample state. (b) Probabilities versus time for $N = 10$ with an opaque sample, as in Fig.~\ref{sketch}~(d). At time $T$, the particle is in the reference state with probability ${\sim} 0.78$ and lost with probability ${\sim} 0.22$. (c) Probability $\ploss$ at time $T$ as a function of $N$ with an opaque sample. The inset shows the same on a double-logarithmic scale.}
\label{rabi}
\end{figure}

%----------------------------------------------------------------
\section{Interaction-free measurements}
\label{ifm}
There are several different schemes to achieve IFM with a high success rate that have been proposed~\cite{Kwiat1995, Putnam2009} or realized~\cite{Kwiat1999, Tsegaye1998} in the literature. Most of them employ techniques based on the quantum Zeno effect, where frequent measurements prevent a quantum system from changing its state~\cite{Misra1977}. It is this kind of ``quantum-Zeno-like'' IFM setup that we will discuss in this article. Example schemes are shown schematically in Fig.~\ref{sketch} for either photons or electrons as probe particles.

We can describe the probe particle in a quantum-Zeno-like IFM setup as a three-state system. The first state is the reference state $\qref$, in which the particle starts out. This state is coupled to a second state, the sample state $\qsam$. In the examples of Fig.~\ref{sketch}, the coupling between the two states is achieved via a beam splitter. The third state of the system is the loss state $\qloss$. It keeps track of the probability that the particle is lost in a sample interaction, for example due to absorption or scattering. Note that $\qloss$ may be a continuum of states.

We start out with the particle in the reference state $\qref$. The coupling strength between $\qref$ and $\qsam$ is such that, in the absence of a sample, the particle is fully in $\qsam$ after $N$ round trips, i.e., $N$ encounters with the beam splitter. This half-finished oscillation from $\qref$ to $\qsam$ takes the time $T$, as shown in Fig.~\ref{rabi}~(a). (A full Rabi oscillation from $\qref$ over $\qsam$ back to $\qref$ would take the time $2T$.) In state $\qsam$, the particle may encounter a sample once during each round trip. As in the quantum Zeno effect, an encounter of the particle with an opaque sample constitutes a measurement of the state of the particle with two possible outcomes: either the particle is still in the reference state, and the oscillation restarts from there, or the particle is in the sample state, where it is subsequently lost. The presence of the sample therefore inhibits the coherent evolution from $\qref$ to $\qsam$, as shown in Fig.~\ref{rabi}~(b) for the example of $N = 10$. To model the probability of losing the particle, we simply transfer the amplitude from $\qsam$ to $\qloss$ every time the particle encounters an opaque sample. 

At time $T$, the observer measures the state of the particle and determines whether it is in $\qref$, $\qsam$, or $\qloss$. This way, the following information about the sample is obtained: if the particle is still in state $\qref$ or if it is lost in $\qloss$, there must be an opaque sample blocking the evolution to $\qsam$. If the particle is found in $\qsam$ at time $T$, the sample is transparent (or there is no sample). An interaction-free measurement is successful if the presence of the sample is detected without loss via a probe particle in $\qref$. To achieve a high success rate, the probability of finding the particle in $\qloss$ needs to be minimized. The loss probability $\ploss$ depends on the number of round trips $N$, and is given by~\cite{Kwiat1995}
\begin{equation}
 \ploss = 1 - \cos^{2N}{\left(\frac{\pi}{2N}\right)}
 \xrightarrow{\mathrm{large}\>N} \frac{\pi^2}{4N},
 \label{eqploss}
\end{equation}
which converges to $0$ for $N \to \infty$, see Fig.~\ref{rabi}~(c). Thus, in principle, the existence of a sample can be ascertained without loss. For an intuitive picture of why more sample encounters ultimately lead to less loss in an IFM, consider the following: while doubling $N$ doubles the number of sample encounters, the loss probability during each interaction is reduced by a factor of $4$ because the coherent build-up of probability in $\qsam$ is approximately quadratic.

Note that the calculations shown in Fig.~\ref{rabi} assume a continuous coupling between $\qref$ and $\qsam$, which occurs in some proposals for IFM~\cite{Putnam2009}, while other proposals and the scheme in Fig.~\ref{sketch} work with one discrete coupling step in every round trip~\cite{Kwiat1995, Kwiat1999}. Instead of the continuous oscillation of probabilities shown in Fig.~\ref{rabi}~(a), the oscillation proceeds in discrete steps for a discrete coupling. These different couplings lead to the same results in this paper because the particle-sample interaction also occurs in discrete steps.

%----------------------------------------------------------------
\section{Semitransparent samples}
\label{alpharesults}

\begin{figure}[t]
\includegraphics[width=\columnwidth]{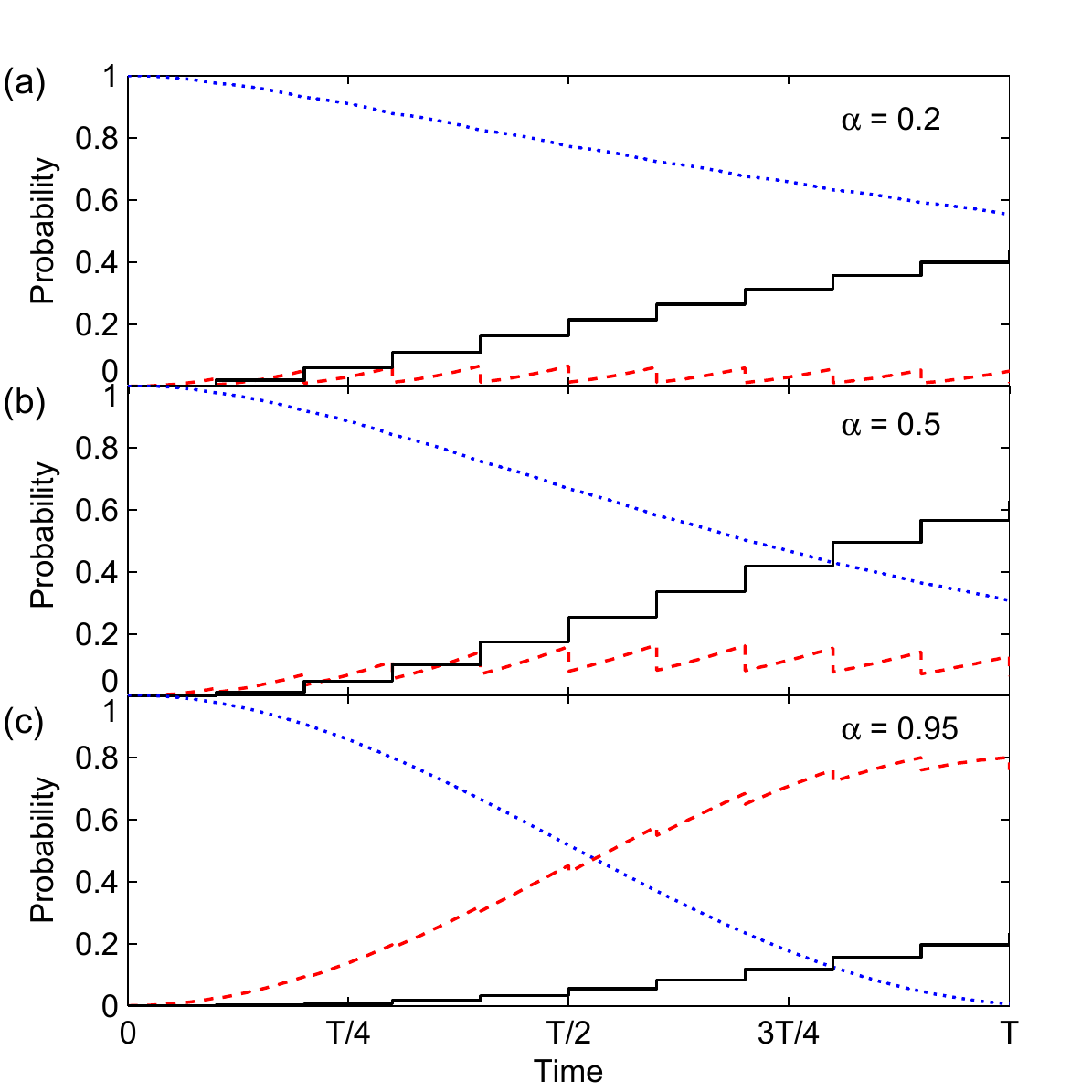}
\caption{Time evolution of the probe particle state probabilities for different transparencies: $\alpha = 0.2$ (a), $\alpha = 0.5$ (b), $\alpha=0.95$ (c). Shown are the probabilities $\pref$ (dotted blue line), $\psam$ (dashed red line), and $\ploss$ (black line) for $N = 10$.}
\label{3alphas}
\end{figure}

So far, we have considered only fully opaque samples, which completely block the coherent build-up of the probe particle wave function in the sample state. This ``all-or-nothing'' IFM scheme is fully loss-free in the limit $N \to \infty$ and thus outperforms any classical detection in terms of sample damage. We now investigate the effect of semitransparent samples in quantum-Zeno-like IFM setups, which are more reminiscent of real-world samples relevant for imaging, e.g., biological samples.

If a particle encounters a semitransparent sample with a transparency $\alpha$, it passes the sample with probability $\alpha$ or it is lost with probability $1-\alpha$. Lost (i.e., absorbed or scattered) particles are the cause of damage during imaging. Additionally, the sample may cause a phase shift $\phi$ of the particle's wavefunction. One encounter with the semitransparent sample leads to the following modification of the particle's amplitudes $r$, $s$, and $l$ in the three states $\qref$, $\qsam$, and $\qloss$:
\begin{equation}
\begin{split}
 \Ket{\psi} & = r \qref + s \qsam + l \qloss \\
 &\to r \qref + e^{i\phi} \sqrt{\alpha} s \qsam + \sqrt{|l|^2 + (1-\alpha)|s|^2} \qloss
% &\to r \qref + e^{i\phi} \sqrt{\alpha} s \qsam + e^{i\phi}(l + \sqrt{1-\alpha}s) \qloss.
\end{split}
\label{sampleint}
\end{equation}
Note that this can be thought of as a combination of a unitary transformation between $\qsam$ and $\qloss$ and a measurement of whether the particle is in $\qloss$ (see also Refs.~\cite{Garcia-Escartin2005} and~\cite{Mitchison2001} for different but equivalent approaches of describing the interaction with a semitransparent sample). The loss from $\qsam$ to $\qloss$ is irreversible and the state $\qloss$ merely serves to `count' the lost amplitude. For $\alpha=0$ we recover the fully opaque case discussed in the last section. Similarly, setting $\alpha = 1$ and $\phi = 0$ represents the case of an absent object.

\begin{figure}[tb]
\includegraphics[width=\columnwidth]{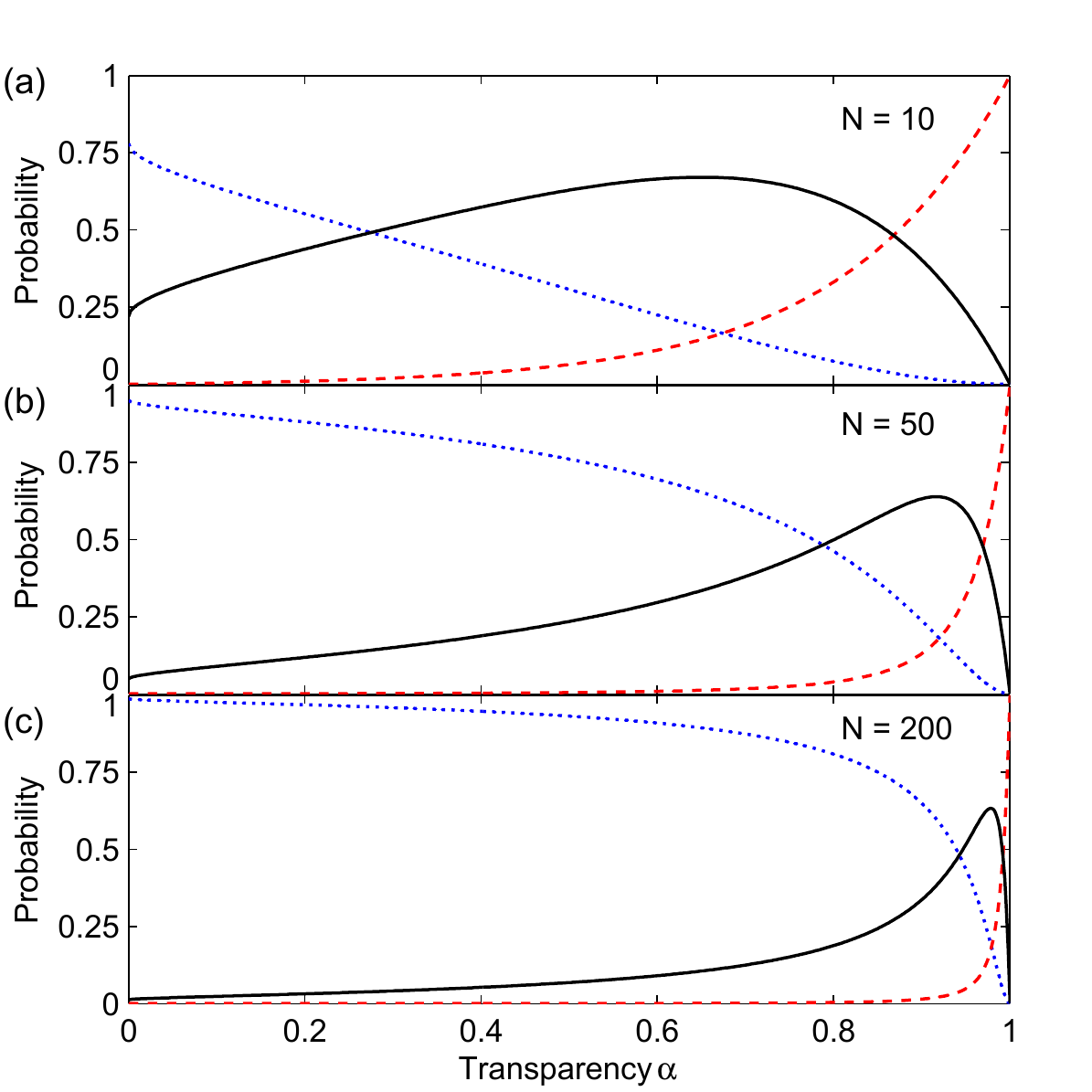}
\caption{Probabilities for the probe particle to be detected in $\qref$, $\qsam$, or $\qloss$ at time $T$ as a function of $\alpha$ for $N=10$ (a), $50$ (b), and $200$ (c). Shown are $\pref$ (dotted blue lines), $\psam$ (dashed red lines), and $\ploss$ (black lines). For low transparencies, the particle is most likely found in the reference state, while the particle is most likely to enter the sample state for high transparencies. In between low and high transparencies, the loss probability has a maximum and the two other probabilities change swiftly.}
\label{results}
\end{figure}

\begin{figure}[tb]
\includegraphics[width=\columnwidth]{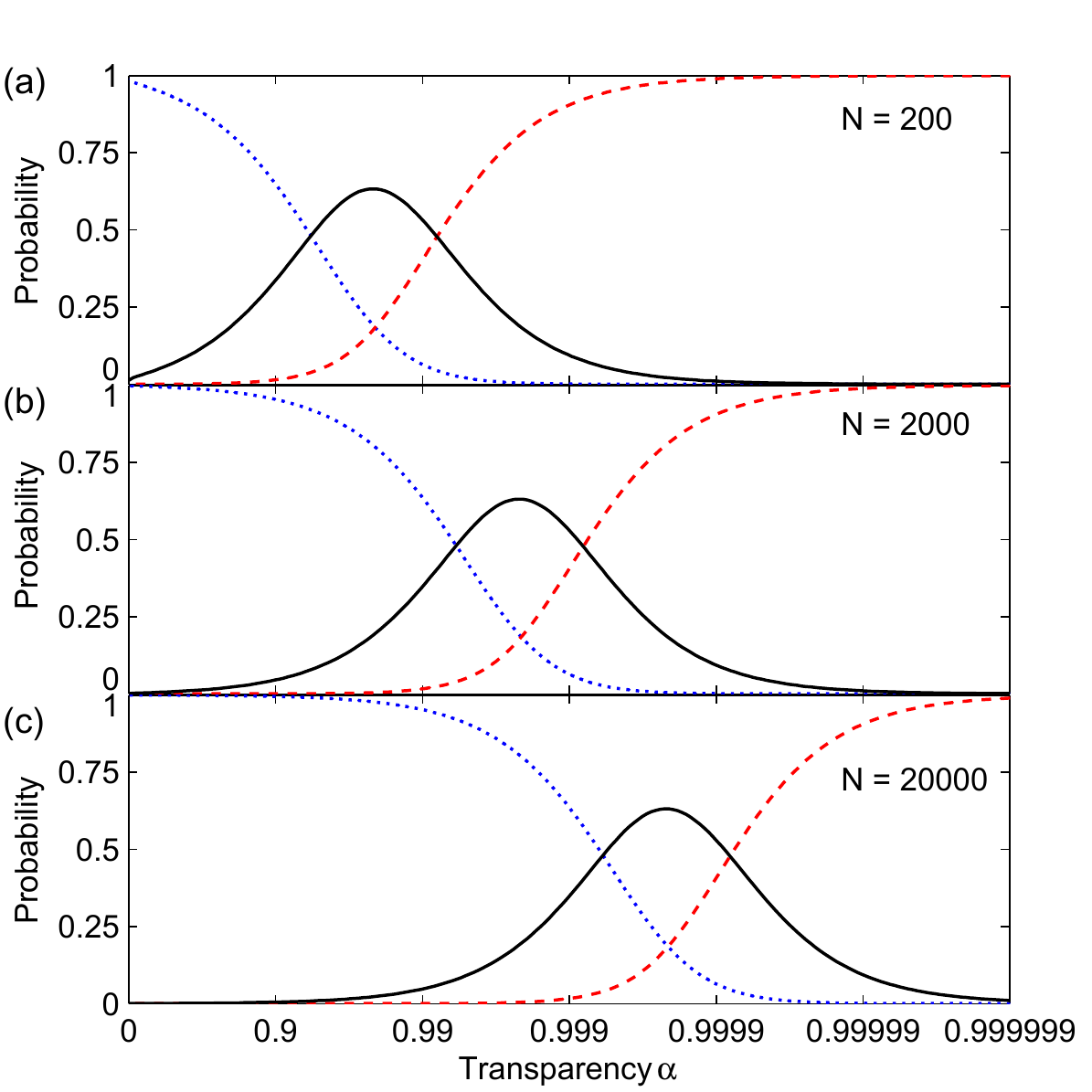}
\caption{Same as Fig.~\ref{results} but for $N=200$ (a), $2000$ (b), and $20000$ (c), and on a logarithmic scale. For large $N$, a change of $N$ simply corresponds to a shift of the probability curves along the logarithmic axis.}
\label{logresults}
\end{figure}

We can now simulate the IFM scheme with arbitrary semitransparent samples by starting a particle in the state $\Ket{\psi} = \qref$ at $t=0$, using a time propagator to simulate the coherent evolution from $\qref$ to $\qsam$, and periodically interrupting this process $N$ times with a sample encounter according to Eq.~\ref{sampleint}. In the $\qref =\bigl(\begin{smallmatrix} 1 \\ 0 \end{smallmatrix} \bigr)$ and $\qsam =\bigl(\begin{smallmatrix} 0 \\ 1 \end{smallmatrix} \bigr)$ basis, the propagator for a coherent evolution over a time interval $\Delta t$ is
\begin{equation}
 \frac{1}{2}
 \begin{pmatrix}
  1 + e^{-i \pi \Delta t / T} & 1 - e^{-i \pi \Delta t / T}\\
  1 - e^{-i \pi \Delta t / T} & 1 + e^{-i \pi \Delta t / T}
 \end{pmatrix}.
\end{equation}
For a time interval of $\Delta t = T/N$, this propagator is equivalent to a beam splitter that completes half an oscillation from $\qref$ to $\qsam$ in $N$ steps. This confirms that the continuous and discrete couplings in different IFM setups lead to the same result, as discussed in the last section.

We now simulate the behavior of a particle in a quantum-Zeno-like IFM setup. We will first discuss the case of a semitransparent sample without a phase shift and will include the phase shift in Sec.~\ref{phaseshift}. The simulation is then defined by two parameters:

\begin{itemize}
 \item the number of round trips $N$ within the duration $T$ of half an oscillation
 \item the transparency $\alpha$ of the sample.
\end{itemize}

Examples of such simulations for different $\alpha$ are given in Fig.~\ref{3alphas}. They show the interplay of coherent build-up from $\qref$ to $\qsam$ and periodic particle loss due to the presence of the semitransparent sample. For an IFM measurement, the relevant results of a simulation are the three probabilities $\pref$, $\psam$, and $\ploss$ at time $T$, when the probe particle state is measured. Fig.~\ref{results} displays these probabilities as a function of $\alpha$ for $N=10$, $50$, and $200$.  $\pref$ starts out close to $1$ for $\alpha=0$ and ends at $0$ for $\alpha=1$, while $\psam$ shows the opposite behavior. The probability of losing the particle is low for $\alpha$ around 0 or 1, but it peaks in between. Both $\pref$ and $\psam$ change swiftly in the region where $\ploss$ peaks.

For low and high transparency, the situation is similar to ``all-or-nothing'' IFM with either a fully transparent or a fully opaque sample: if the transparency is low, the quantum Zeno effect prevents the evolution to the sample state. If the transparency is high, the loss is negligible, so the particle can enter the sample state. For intermediate transparencies, significant parts of the wavefunction can enter the sample state and can be lost. Then there is a high chance of losing the particle. The position of this high-loss region shifts to higher transparencies as the number of passes through the sample increases.

Analyzing the loss peak for different $N$ shows that its height $\max(\ploss)$ slowly decreases for larger $N$, converging to a value of ${\sim} 0.63$, while the position of the peak, $\alpha'$, shifts to high transparencies. The behavior of the probabilities at large $N$ is easier to discern on a logarithmic scale, which is shown in Fig.~\ref{logresults} for $N=200$, $2000$, and $20000$. Here, an increase of $N$ simply corresponds to a shift of the probability curves further towards $\alpha = 1$. This allows us to give an approximate formula for the position of the maximum of the loss peak $\alpha'$ for $N \gg 1$:

\begin{equation}
 \alpha' \approx 1 - \frac{4.4}{N}.
 \label{alphaprime}
\end{equation}

\begin{figure}[t]
 \includegraphics[width=\columnwidth]{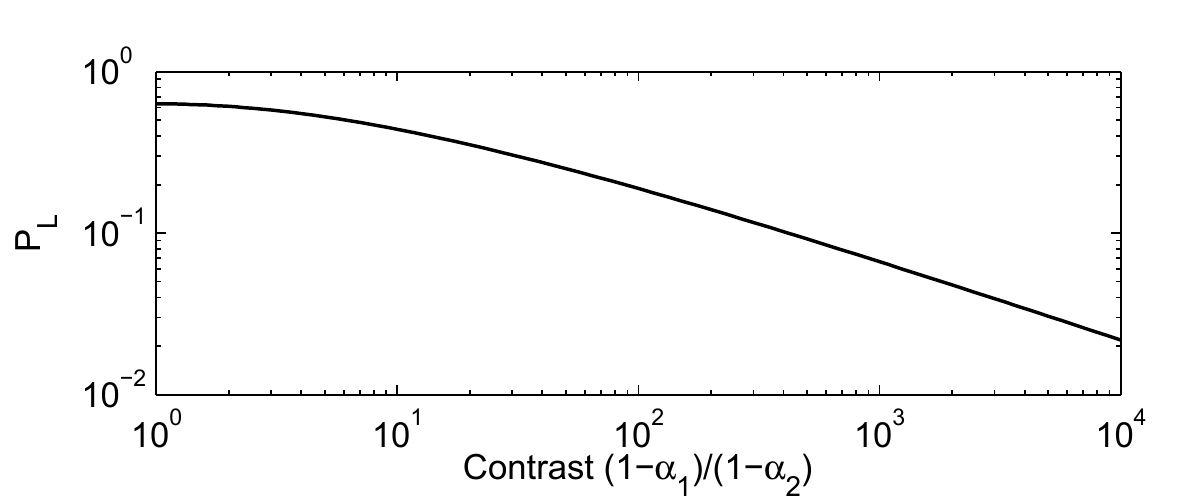}
\caption{Loss probability as a function of the contrast $(1-\alpha_1)\,/\,(1-\alpha_2)$ if two transparencies $\alpha_1$ and $\alpha_2$ are to be distinguished in an IFM with $N \gg 1$, assuming that $N$ is always chosen so as to minimize the average loss probability at $\alpha_1$ and $\alpha_2$.}
\label{contrast}
\end{figure}

We now directly see that the ``all-or-nothing'' IFM process of distinguishing the presence from the absence of an object also works for distinguishing between samples with high transparencies $\alpha_2$ and low transparencies $\alpha_1$ if the contrast is high, i.e., if $(1-\alpha_1)\,/\,(1-\alpha_2) \gg 1$. In this case, $N$ can be chosen so that $\alpha'$ lies between the two transparencies and the loss probability is low for both $\alpha_1$ and $\alpha_2$. Example transparencies are $\alpha_1 = 0.9$ and $\alpha_2 = 0.9999$ for $N=2000$ as in Fig.~\ref{logresults}~(b). Of course, the loss probability of such a measurement is always higher than in an ``all-or-nothing'' IFM with the same $N$: $\ploss$ is determined by the contrast, which can only be high for $\alpha_2$ close to $1$. The loss probability in an IFM as a function of the contrast is plotted in Fig.~\ref{contrast}. This plot assumes that all IFMs are performed with the optimum number of round trips $N$, which minimizes the average loss probability at the given transparencies $\alpha_1$ and $\alpha_2$. An approximate formula for the optimum $N$ can be obtained from Eq.~\ref{alphaprime} and the average on the logarithmic scale (i.e., the geometric mean) of $1 - \alpha_1$ and $1 - \alpha_2$:
\begin{equation}
 \nopt \approx \frac{4.4}{\sqrt{(1-\alpha_1)(1-\alpha_2)}}.
 \label{noptapprox}
\end{equation}
If the contrast is large and $\alpha_1 \gg 0$, $\nopt$ represents a good approximation of the exact optimum $N$, which can be found numerically.

If $\alpha_2=1$, the contrast is infinite and the loss probability can approach $0$ by increasing $N$ and thus bringing $\alpha'$ arbitrarily close to $1$. This implies that interaction-free measurements can detect the presence or absence of any transparent object with arbitrarily low loss. That this is possible was already shown by Azuma~\cite{Azuma2006}. There is also a general finding on quantum measurements by Mitchison and Massar~\cite{Mitchison2001}, which states that it is possible to distinguish an object with any transparency $\alpha_1<1$ from a perfectly transparent object $\alpha_2 = 1$ without loss, but it is not possible to distinguish without loss between two semitransparent objects with transparencies $\alpha_{1,2} < 1$. The concrete cases shown in this section will be important for the discussion that follows.

%----------------------------------------------------------------
\section{Distinguishing two transparencies}
\label{twotra}

\begin{figure*}[bt]
% \begin{figure}[p]
\begin{center}
 \includegraphics[width=.48\textwidth]{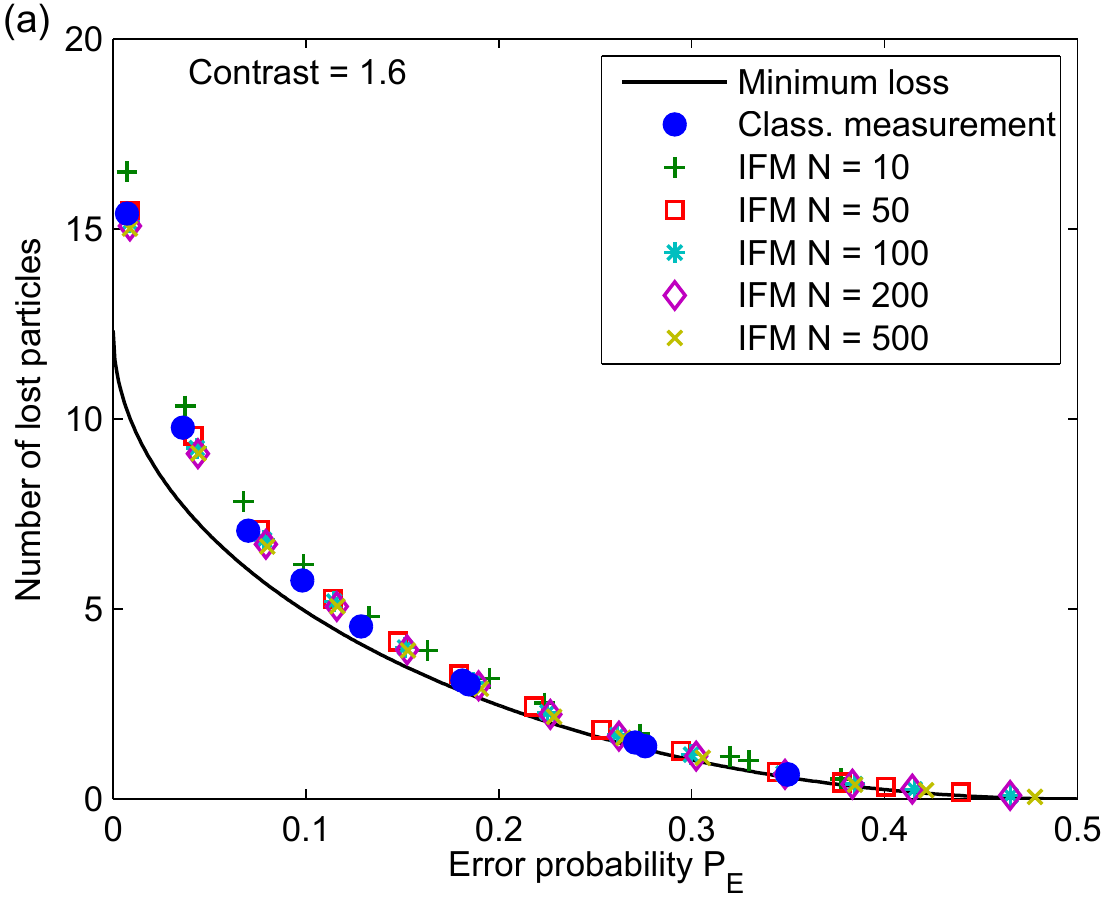}
 \includegraphics[width=.48\textwidth]{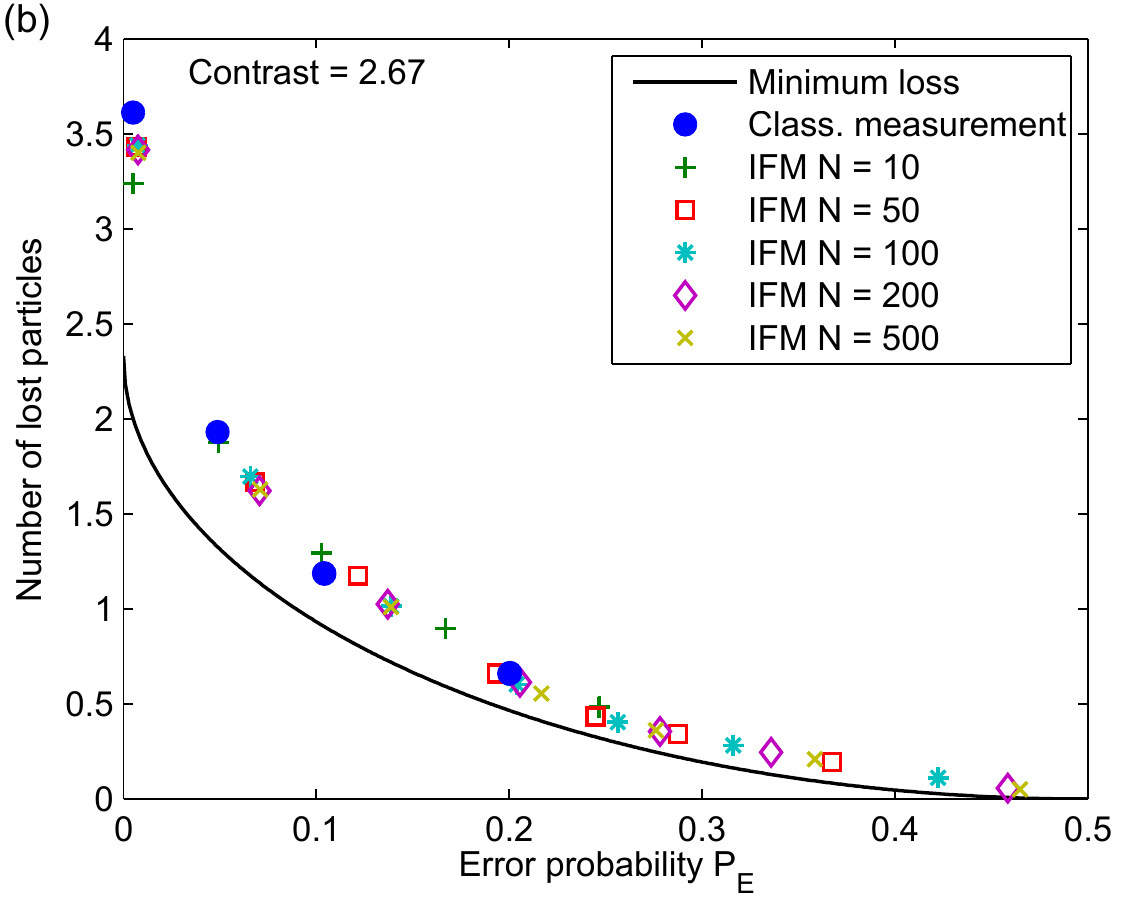}\\
 \includegraphics[width=.48\textwidth]{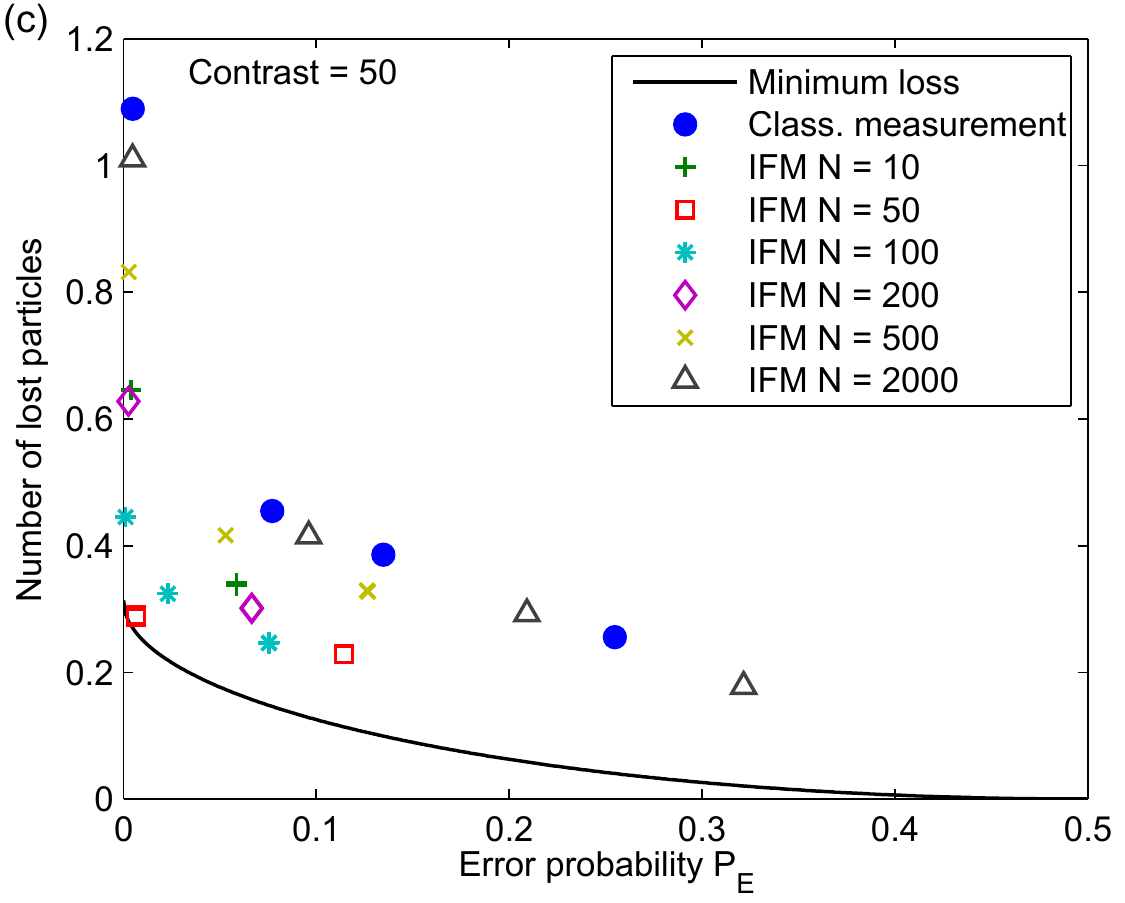}
 \includegraphics[width=.48\textwidth]{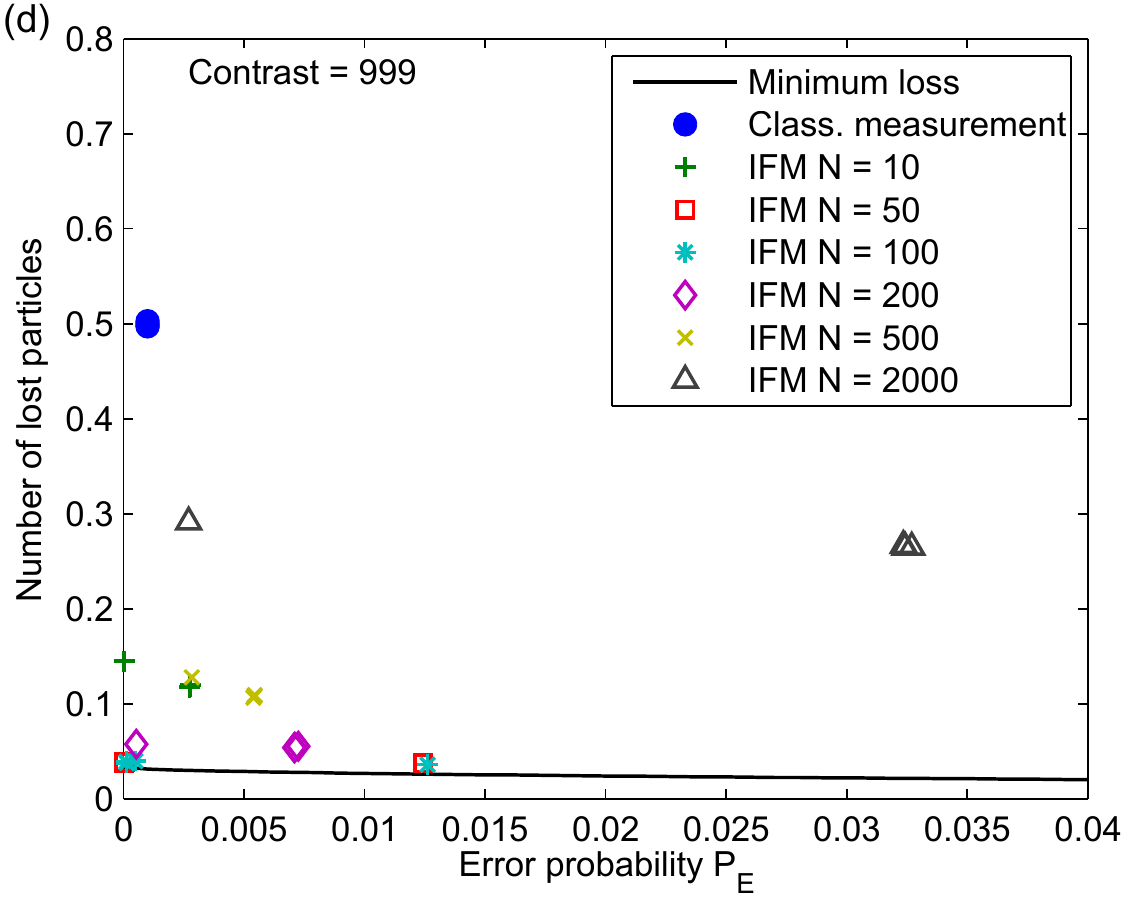}
\caption{Relationship between the average number of lost particles and the error probability in a measurement to distinguish two transparencies. Shown here are results from simulations of a classical transmission measurement (blue circles), IFM with various parameters (other symbols), and the minimum amount of loss in such a measurement according to Eq.~\ref{mitchi} (black line). The task is to distinguish the two transparencies (a) $\alpha_1 = 0.2$ and $\alpha_2 = 0.5$, (b) $\alpha_1 = 0.04$ and $\alpha_2 = 0.64$, as in Fig.~3 of Ref.~\cite{Mitchison2002}, (c) $\alpha_1 = 0.5$ and $\alpha_2 = 0.99$, (d) $\alpha_1 = 0.001$ and $\alpha_2 = 0.999$. Each data point was obtained by $40000$ simulation runs. For example, to distinguish the two transparencies in (c) a classical measurement loses ${\sim}0.45$ particles on average to reach an error probability of ${\sim}0.8$, while an IFM with $N = 100$ can reach the same error probability with only ${\sim}0.25$ lost particles and the minimum amount of loss for this error probability is ${\sim}0.15$.}
\label{twoalpha}
\end{center}
\end{figure*}

In this section, we discuss the task of distinguishing two a priori given transparencies $\alpha_1$ and $\alpha_2$. We simulate how many particles are lost on average as a function of the measurement error probability. We compare the results of a quantum-Zeno-like IFM scheme to classical measurements and to the minimum number of lost particles in a general quantum measurement~\cite{Mitchison2002}.

In a classical measurement, one possibility is to simply count how many particles are transmitted through the sample, i.e., have not been absorbed or scattered out of the beam. In an IFM, one can count how many particles are detected in either the reference state, the sample state, or the loss state. The counts can then be compared to the three probabilities $\pref$, $\psam$, and $\ploss$ to infer which of the two given transparencies is more likely.

To reduce the error probability $P_E$ of wrongly identifying the sample, we need to increase the number of probe particles. A measurement is therefore always a trade-off between the number of lost particles, which determines the sample damage, and the error probability.

We evaluate both classical and interaction-free measurements using Monte Carlo simulations, similar to the analysis of classical measurements in Ref.~\cite{Mitchison2002}. One measurement to distinguish transparencies consists of multiple runs with single probe particles. After each run, the conditional probabilities of the transparency being either $\alpha_1$ or $\alpha_2$ given the current measurement result is evaluated. If either of the probabilities is below a chosen threshold $x$, the measurement is stopped with the more likely $\alpha$ as the result. Otherwise, the measurement continues for another run.

For a classical transmission measurement after $n$ runs with $n'$ detected particles, i.e., $n'$ particles that traversed the sample without being lost, the conditional probability of $\alpha=\alpha_1$ is
\begin{equation}
 P(\alpha=\alpha_1) = \frac{\alpha_1^{n'} (1-\alpha_1)^{n-n'}}
 {\alpha_1^{n'} (1-\alpha_1)^{n-n'} + \alpha_2^{n'} (1-\alpha_2)^{n-n'}}
\end{equation}
while $n-n'$ particles are lost. For an IFM after $n$ runs with $n_r$ particles detected in the reference state, $n_s$ particles detected in the sample state, and $n_l$ particles lost, the equivalent probability is:
\begin{widetext}
\begin{equation}
 P(\alpha=\alpha_1) = \frac{\pref(\alpha_1)^{n_r} \psam(\alpha_1)^{n_s} \ploss(\alpha_1)^{n_l}}
 {\pref(\alpha_1)^{n_r} \psam(\alpha_1)^{n_s} \ploss(\alpha_1)^{n_l} + 
 \pref(\alpha_2)^{n_r} \psam(\alpha_2)^{n_s} \ploss(\alpha_2)^{n_l}}.
\end{equation}
\end{widetext}

We use Monte Carlo simulations to find out the average number of lost particles as well as the average error that arises during the classical and the IFM measurement schemes. The relationship between error probability and loss is found by varying the error threshold $x$.  Results for the number of lost particles versus error probability are shown in Fig.~\ref{twoalpha}. Note that, while the error threshold can be varied continuously, the measurement scheme is ultimately discrete, so different error thresholds may lead to the same result. For this reason, the relationship between lost particles and error probability cannot be given as a continuous function, but only on discrete points. Especially if $\alpha_1$ and $\alpha_2$ have a high contrast, the number of discrete points is quite small as only a few particles are required for distinguishing the transparencies with low error probability.

Fig.~\ref{twoalpha}~(a) and (b) show a measurement of two transparencies with low contrast, while panels (c) and (d) show a measurement of two transparencies with high contrast. In the low-contrast case, the results for IFM do not depend much on $N$ and are usually similar to a classical transmission measurement. In this case, IFM does not offer a benefit over classical measurements. This is true for any combination of low-contrast transparencies we tried.

IFM outperforms the classical measurement for high contrasts, as shown in Fig.~\ref{twoalpha}~(c) and (d). Here, more particles are lost in a classical measurement than in an IFM, and there is a clear difference between IFMs with different $N$. In (c), the optimum $N$ with the best ratio between lost particles and low error is $N=50$, while $N=50$ and $N=100$ perform approximately equally well in (d), where both achieve a smaller error probability than classical measurements with an order of magnitude less particles lost. At optimum $N$ the maximum of the loss probability is in between $\alpha_1$ and $\alpha_2$ so $\ploss$ is small for both, as discussed in the last section. Calculating $\nopt$ numerically, we obtain $N=54$ for (c) and $N=73$ for (d), in good agreement with the simulation results.

For $\alpha_{2}$ closer to $1$, the number of lost particles will decrease further while the optimum $N$ will increase. Note that only a single particle is often already sufficient to distinguish the two transparencies with a low error probability in a high-contrast IFM as in Fig.~\ref{twoalpha}~(d). In this case, the average number of lost particles is directly determined by the loss probability given in Fig.~\ref{contrast}.

We now compare our results for both classical and IFM to the minimum number of lost particles in a quantum measurement, as derived in Ref.~\cite{Mitchison2002}, which is valid for a more general quantum measurement scheme than the IFM setup discussed here. While the minimum number of lost particles can be reached using a non-constant coupling between $\qref$ and $\qsam$ that has to be adapted to the given transparencies $\alpha_1$ and $\alpha_2$, the quantum-Zeno-like IFM setup discussed here is technologically less challenging and has already been realized experimentally using photons as probe particles~\cite{Kwiat1999}.
 
If a sample can have two transparencies $\alpha_1$ or $\alpha_2$ with equal probability and these transparencies are to be distinguished with an error probability of at most $P_E$, then the minimum number of lost particles is
\begin{equation}
 \nmin = \frac{\sqrt{1-\alpha_1}\sqrt{1-\alpha_2}\left(1-2\sqrt{P_E(1-P_E)}\right)}{1-\sqrt{\alpha_1 \alpha_2} - \sqrt{1-\alpha_1}\sqrt{1-\alpha_2}},
 \label{mitchi}
\end{equation}
following Eq.~(1) of Ref.~\cite{Mitchison2002}. Eq.~\ref{mitchi} is plotted together with the simulation results in Fig.~\ref{twoalpha}. Using the optimum number of round trips, quantum-Zeno-like IFM can approach the minimum number of lost particles if the contrast of $\alpha_{1,2}$ is high. In the case of low contrast, the minimum number of lost particles is somewhat smaller than what is achieved in either classical or IFM measurements, but leaves room for only a factor of $2$ improvement for the error probabilities considered here.

%----------------------------------------------------------------
\section{Measuring an unknown transparency}
\label{onealpha}
Arguably the most common task in imaging is to measure an unknown transparency $\alpha$. We determine the number of lost particles in an IFM in this case and compare it to a classical measurement.

\begin{figure*}[tb]
\begin{center}
 \includegraphics[width=.4\textwidth]{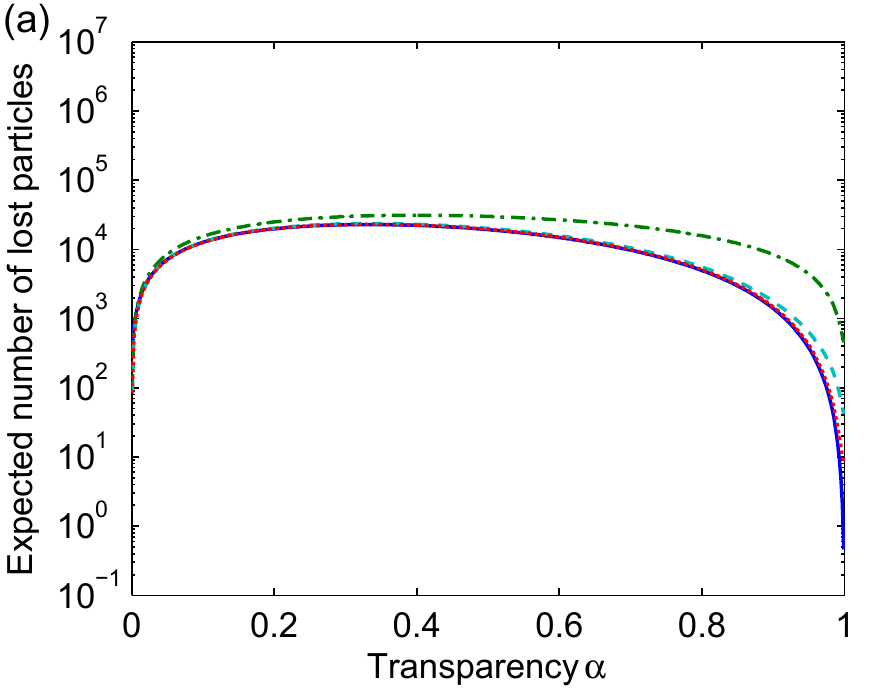}
 \hspace{.5cm}
 \includegraphics[width=.4\textwidth]{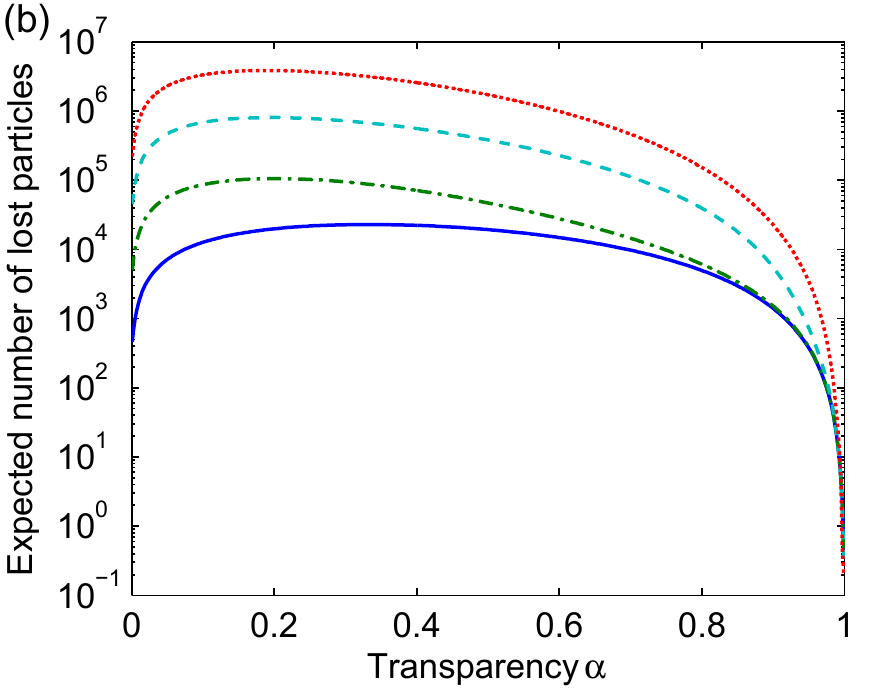}
\caption{Expected number of lost particles in a measurement of $\alpha$ with uncertainty $\dal \leq 0.01$ as a function of transparency using the reference state (a) or the sample state (b) as signal. The curves for a classical measurement (blue line), and IFM with $N=10$ (dash-dotted green line), $N=100$ (dashed cyan line), and $N=500$ (dotted red line).}
\label{clop}
\end{center}
\end{figure*}

As in the previous section, running the IFM process $M$ times with a semitransparent sample results in a multinomial distribution of particles that are either detected in $\qref$, detected in $\qsam$, or lost in $\qloss$. Here, $M$ is the number of times the IFM is repeated, as opposed to $N$, which denotes the number of round trips in a single IFM. The number of particles detected in these states can be used to estimate the probabilities $\pref(\alpha)$, $\psam(\alpha)$, and $\ploss(\alpha)$ (see Fig.~\ref{results}), which yield the transparency $\alpha$.

We assume that we have no prior knowledge about the transparency $\alpha \in [0,1]$ of a sample. We consider the number of particles detected either in the reference state or in the sample state independently. The number of particles found in one of these states after running the process $M$ times follows a binomial distribution with probability $P(\alpha) = \pref(\alpha)$ or $\psam(\alpha)$, respectively. After estimating the probability $P(\alpha)$ from such a measurement, the uncertainty of the measurement $\Delta P$ (i.e., the confidence interval) can be estimated from the normal approximation of the binomial distribution. For a $95\,\%$ confidence level this yields
\begin{equation}
 \dpr \approx (2 \cdot 1.96) \sqrt{\frac{P(\alpha)(1-P(\alpha))}{M}},
 \label{normapprox}
\end{equation}
where the factor $2 \cdot 1.96$ is due to the $97.5$ percentile point of the normal distribution. To obtain the uncertainty in transparency $\dal$ from the uncertainty $\dpr$, we use the relation $\dpr = P'(\alpha) \dal$, which is valid if the slope $P'(\alpha)$ does not change significantly within the interval $\dal$. For any transparency, Eq.~\ref{normapprox} can be inverted to find the number of trials necessary for obtaining $\alpha$ with a given uncertainty. Finally, by multiplying this number with the loss probability $\ploss(\alpha)$, one obtains the expected number of lost particles $\nloss$ during a measurement of $\alpha$ with a given uncertainty $\dal$:
\begin{equation}
 \nloss \approx \ploss(\alpha) P(\alpha)(1-P(\alpha)) \left(\frac{3.92}{\dal P'(\alpha)}\right)^2.
 \label{normapprox2}
\end{equation}
We see that $\nloss$ depends on three factors: the loss probability $\ploss(\alpha)$ and the slope and value of the signal $P(\alpha)$. This is true for both IFMs, where we obtain the probabilities from simulations, and classical measurements, where the probabilities are simply $P(\alpha) = \alpha$ and $\ploss(\alpha) = 1-\alpha$.

While the normal approximation allows an estimation of the number of lost particles via a simple analytic formula, it is not valid for $P$ close to $0$ or $1$, where the estimated uncertainty goes to $0$. In the calculations discussed below, we therefore use a different method to find the uncertainty $\dpr$: the Clopper-Pearson confidence interval~\cite{Clopper1934}. This confidence interval is considered conservative, i.e., it tends to overestimate the uncertainty, but it performs significantly better for $\alpha$ close to $0$ or $1$ than the normal approximation~\cite{Brown2001}.

We now compare the number of lost particles in IFM with various $N$ and in a classical transmission measurement. For the comparison, we choose a desired uncertainty $\dal$. We convert this to a probability uncertainty $\dpr$. For any $\alpha$, we then invert the Clopper-Pearson method (with a coverage of $95\,\%$) numerically to find the minimum number of trials $M$ needed to get a confidence interval smaller than $\dpr$. Finally, we multiply $M$ with $\ploss$ to obtain the number of lost particles.

\begin{figure*}[bt]
\begin{center}
 \includegraphics[width=.4\textwidth]{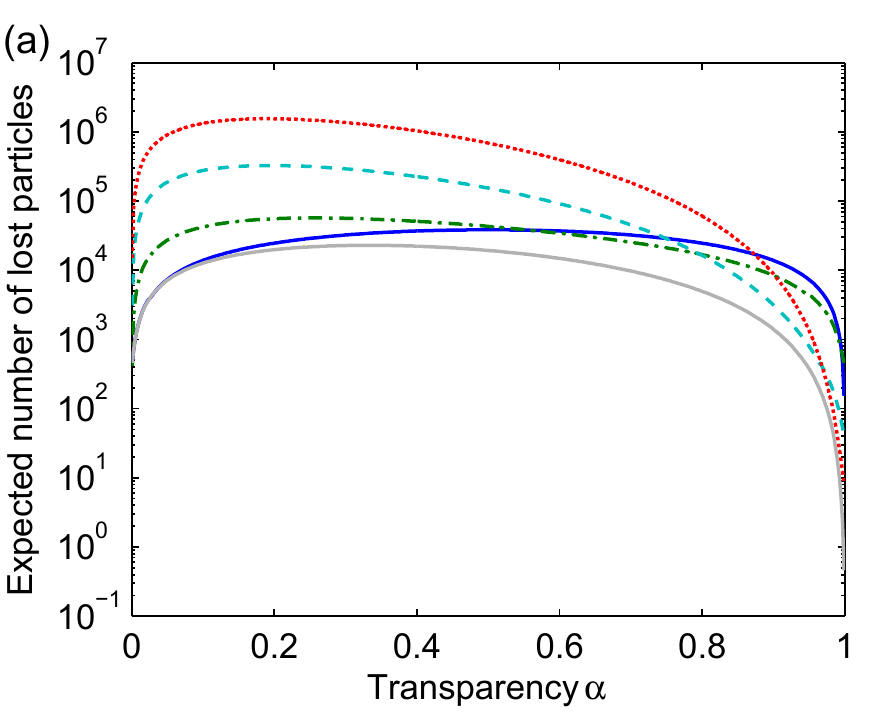}
 \hspace{.5cm}
 \includegraphics[width=.4\textwidth]{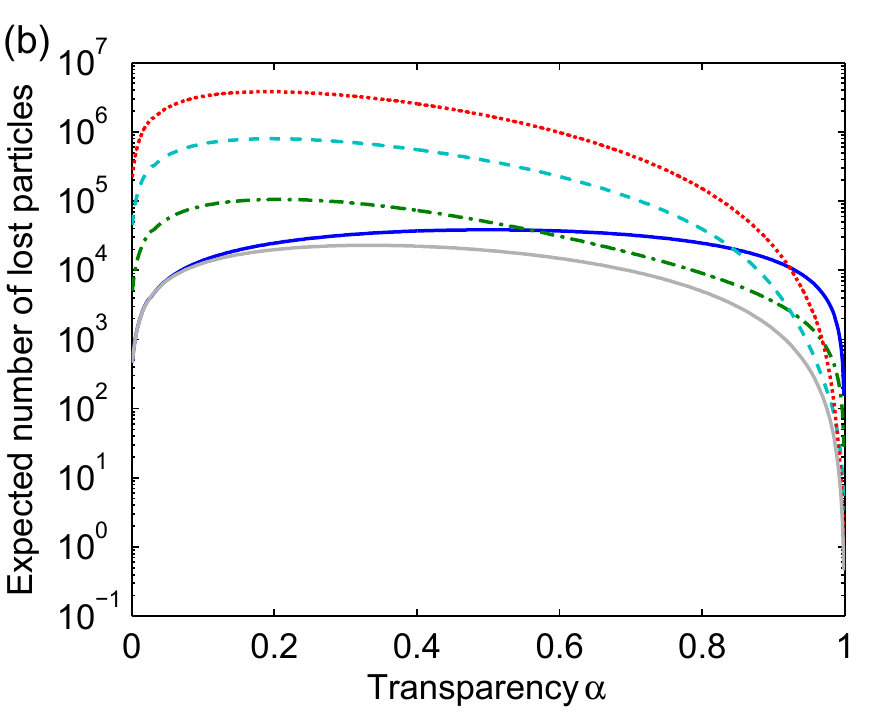}
\caption{Expected number of lost particles in a measurement of $\alpha$ with uncertainty $\dal \leq 0.01$ as a function of transparency using the reference state (a) or the sample state (b) as signal and using Poissonian statistics. The curves shown here are for a classical measurement (blue line), and IFM with $N=10$ (dash-dotted green line), $N=100$ (dashed cyan line), and $N=500$ (dotted red line). For comparison to binomial statistics, the result of a binomial classical measurement is shown as a light gray line (same as the blue solid line in Fig.~\ref{clop}). The binomial measurement outperforms all Poissonian measurements in terms of lost particles.}
\label{poisson}
\end{center}
\end{figure*}

Results for an uncertainty $\dal = 0.01$ are shown in Fig.~\ref{clop}. Note that the relative behavior of the curves does not significantly depend on the value of $\dal$, while the absolute number of lost particles increases quadratically with $\dal^{-1}$.

The results show, first of all, that using the number of particles in the sample state as signal always results in more damage than a classical transmission measurement. This is because the signal slope is very small for small transparencies, while classical measurements have a lower damage probability at high transparencies.

Using the reference state as signal gives a more interesting result: while more particles are lost for low $N$, the number of lost particles is almost exactly the same as for a classical measurement for all $N \gtrsim 50$. This is somewhat unexpected as the signal curves are significantly different for different $N$ (see Figs.~\ref{results} and \ref{logresults}). However, we find that the changes of slope, signal, and loss probability compensate each other, so the overall number of lost particles is the same. This phenomenon is also visible in Fig.~\ref{twoalpha}~(a) and (b) in the last section, where classical and interaction-free measurements also perform very similarly.

Note that it would be possible, in principle, to combine information from the reference and sample signal to achieve lower damage. However, as many more particles are required to gain information from the sample signal than from the reference signal, the final result would be similar to the result using only the reference state shown in Fig.~\ref{clop}~(a). Another possibility is using the number of lost particles as signal if it can be measured. Like the reference and sample signals, however, using the loss as signal also does not lead to lower loss than a classical measurement.

We conclude that IFM does not offer a benefit over classical measurements for determining an unknown transparency precisely.

\section{Poissonian statistics}
\label{poissstat}
So far, we assumed that the particles in the IFM process can be sent in one after another and the total number of particles is known. This led to a detection process governed by binomial or multinomial statistics. Most sources of particles that may be used for imaging or IFM, however, do not produce such number states of particles. Instead, only the average number of particles $\mean{M}$ in any given amount of time is typically known, while the actual number $M$ follows Poissonian statistics. We will now discuss how this affects both classical and IFM measurements of semitransparency.

If we consider a classical measurement with a Poissonian source of particles, we need to distinguish two cases: whether the number of lost particles can be detected or whether it is unknown. For example, lost electrons in electron microscopy are usually scattered by the sample, so the number of lost particles can be measured, while lost photons in light microscopy are often absorbed, so the number of lost particles is unknown. Both the number of detected and the number of lost particles obey a Poisson distribution with average number $\alpha \mean{M}$ and $(1-\alpha) \mean{M}$, respectively. Similarly, the numbers of detected and lost particles in an IFM scheme also follow Poissonian statistics with average numbers $\pref \mean{M}$, $\psam \mean{M}$, and $\ploss \mean{M}$. If lost particles can be measured and the information from all measurements is combined, the error and number of lost particles in Poissonian and binomial statistics are approximately identical. This is because the number of all particles is counted in this case.

Significant differences only occur if the number of lost particles is not accessible. In this case, the normal approximation of the Poisson distribution yields the expected number of lost particles, similar to the previous section:
\begin{equation}
 \nloss \approx \ploss(\alpha) P(\alpha) \left(\frac{3.92}{\dal P'(\alpha)}\right)^2.
\end{equation}
The only difference to the binomial case given in Eq.~\ref{normapprox2} is the dependence on $P(\alpha)$ here vs.\ $P(\alpha)(1-P(\alpha))$ there. This shows that the Poisson distribution behaves similarly to the binomial distribution for small signals $P(\alpha)$ but loses more particles for large signals. As we will see below, this feature of the Poisson distribution leads to a large amount of loss for classical measurements of high transparencies, which allows IFMs to outperform classical measurements there.

We now compare the expected number of lost particles in classical measurements and IFMs for different transparencies. We do not use the normal approximation in the calculation because it is invalid for $P(\alpha)$ close to $0$. Instead, we use the $\chi^2$ method to obtain the confidence intervals. Apart from this, the calculation is identical to the Clopper-Pearson calculation in the previous section. The results are shown in Fig.~\ref{poisson}, which shows the expected number of lost particles in classical and interaction-free measurements of an unknown transparency $\alpha$. It is the same as Fig.~\ref{clop} except for Poissonian statistics instead of binomial statistics, assuming that the number of lost particles \emph{cannot} be measured.

First of all, we find that the amount of lost particles using Poissonian statistics is always higher than using binomial statistics. For $\alpha \gtrsim 1/2$, we find that IFMs achieve the same level of uncertainty as classical measurements with fewer lost particles. For very high transparencies $\alpha \gtrsim 0.95$, they can cause over an order of magnitude less loss. For low transparencies $\alpha \lesssim 1/2$ on the other hand, classical measurements show the least amount of loss.

Similar results hold in the comparison of two transparencies. If two transparencies $\alpha_{1,2} \gtrsim 1/2$ are to be distinguished, IFM schemes perform better than classical measurements even if the contrast is not large. For high contrast, the great reduction of loss discussed in section~\ref{twotra} also works with Poisson distributions.

We conclude that, in addition to the case of high contrast, interaction-free measurements also outperform classical measurements for a particle source with Poissonian statistics if $\alpha \gtrsim 1/2$ and if the number of lost particles cannot be detected.

%----------------------------------------------------------------
\section{Phase shifts}
\label{phaseshift}

So far, we have discussed semitransparent samples without any phase shift of the probe particle. In this section, we assume a fully transparent sample $\alpha=1$, which induces a phase shift $\phi$ during every round trip. After $N$ round trips, the particle can be either in the reference state $\qref$ or in the sample state $\qsam$. 

\begin{figure}[t]
\includegraphics[width=\columnwidth]{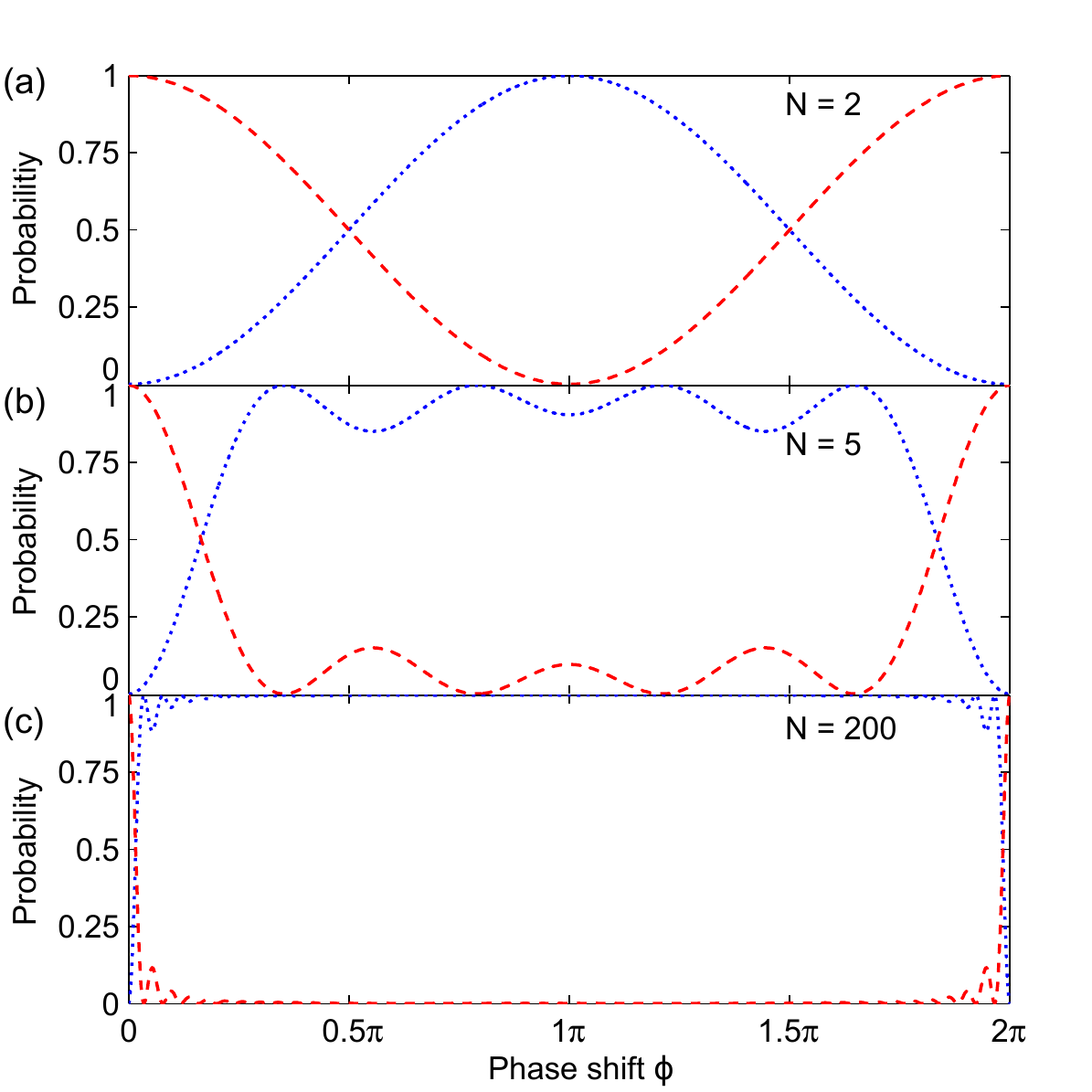}
\caption{Results for an IFM process of a fully transparent sample as a function of the phase shift $\phi$ induced by the sample for $N=2$ (a), $N=5$ (b), and $N=50$ (c). Shown here are $\pref$ (dotted blue line) and $\psam$ (dashed red line).}
\label{deltaphi}
\end{figure}

Results for the probabilities $\pref$ and $\psam$ at time $T$ for different $N$ are shown in Fig.~\ref{deltaphi}. For $\phi = 0$ or a multiple of $2\pi$, the particle is always found in $\qsam$ after $T$. If $N=2$, we see that the probabilities are simply $\psam = \cos^2 (\phi/2)$ and $\pref = \sin^2 (\phi/2)$. This configuration is equivalent to a Mach-Zehnder interferometer. If $N$ becomes larger, the probability is transferred from $\qsam$ to $\qref$ for smaller phase shifts. For large $N$, almost any phase shift leads to the particle being found in $\qref$ after $T$. A possible application of this behavior may be the detection of small phase shifts.

In an IFM measurement with large $N$, a transparent sample with a phase shift $\phi$ will most likely appear the same as an opaque sample. The phase shift leads to a dephasing of sample state and reference state and thus prevents the coherent transfer of amplitude from $\qref$ to $\qsam$. In general, IFM does not allow for the distinction between whether the sample induces a phase shift or whether the sample is opaque. However, if the phase shift is known, it can be compensated with an opposite phase shift. So if two transparencies with a high contrast are to be distinguished and the phase shift induced by the sample with high transparency is known, IFM can still be used by applying an inverse phase shift to the sample state in the IFM setup.

%----------------------------------------------------------------
\section{Conclusion}
In conclusion, interaction-free measurements outperform classical measurements of transparency in special cases. First, we find that IFMs achieve lower loss than classical measurements when samples with a high contrast are to be imaged, which is an approximation of the standard ``all-or-nothing'' IFM and works the same way. Thus, IFMs may significantly reduce the sample damage in imaging of high-contrast samples. Note that the contrast of samples can be artificially enhanced. In transmission electron microscopy for example, staining a sample with metal nanoparticles may be used to achieve a high contrast~\cite{Hayat2000}.

A second regime where interaction-free measurements achieve the same result as classical measurements with fewer lost particles is when the number of particles sent in exhibits a Poisson distribution, the number of lost particles cannot be measured, and the transparency is greater than approximately $1/2$. In this case, the advantage of IFMs is not due to an exploitation of the quantum Zeno effect but due to statistical properties of the Poisson distribution.

Finally, we have shown that phase shifts of a sample have to be compensated for in order to measure the transparency of the sample in an IFM. This is because a sample with a non-zero phase shift may appear the same as an opaque sample. Conversely, the sensitivity of IFMs to phase shifts may be exploited to detect small phase shifts in transparent samples.
\clearpage

\section*{Acknowledgments}
We thank all the members of the Quantum Electron Microscope collaboration for excellent scientific discussions. This research is funded by the Gordon and Betty Moore Foundation.

\end{document}